\documentclass[usegraphicx,usenatbib,letterpaper]{emulateapj}

\usepackage{ulem}
\usepackage{color}
\usepackage{graphics,graphicx}
\usepackage{times} 
\usepackage{amssymb}
\usepackage{amsmath}
\usepackage{natbib}
\usepackage{url}
\usepackage{multirow}
\usepackage{ctable}
\usepackage{threeparttable}
\newif\ifAMStwofonts
\AMStwofontstrue

\def\xmm{{\it XMM-Newton}}

\def\suzaku{{\it Suzaku}}

\def\epicmos1{{\it EPIC}{\rm-MOS1~\/}}
\def\epicmos2{{\it EPIC}{\rm-MOS2 ~\/}}
\def\epicmos{{\it EPIC}{\rm-MOS}}
\def\xis{{\rm XIS}}

\def\nustar{{\it NuSTAR}}

\def\kmps{\hbox{$\rm\thinspace km~s^{-1}$}}

\def\H0{{\rm ~km~s^{-1}~Mpc^{-1}}}

\def\kev{\hbox{\rm keV}}

\def\atpcm{{\rm atom~cm$^{-2}$}}

\def\ergps{\hbox{erg~s$^{-1}$}}

\def\msun{\hbox{$\rm M_{\odot}$}}

\def\Mdot{$\dot{M}$}
\def\mstar{$M_{\rm *}$}

\def\chisq{{$\chi^{2}$}}
\def\rchi{{$\chi^{2}_{\nu}$~\/}}

\def\xspec{\hbox{\small XSPEC}}

\def\heasoft{\hbox{\rm{\small HEASOFT}}}
\def\xselect{\hbox{\rm{\small XSELECT}}}

\def\ftool{\hbox{\rm{\small FTOOL}}}

\def\addascaspec{\hbox{\rm{\small ADDASCASPEC~\/}}}

\def\addascaspec{\hbox{\rm{\small ADDASCASPEC}}}

\def\grid25{\hbox{\rm{\small GRID25}}}

\def\tbnew{\rm{\small TBNEW}}

\def\comptt{\rm{\small COMPTT}}

\def\heii{\rm He\,{\small II}}
\def\ka{K\,$\alpha$}

\def\zsun{$Z_{\odot}$}

\def\etal{et al.~\/}
\def\eg{{\it e.g.~\/}}

\def\ie{{\it i.e.~\/}}

\def\la{\mathrel{\hbox{\rlap{\hbox{\lower4pt\hbox{$\sim$}}}{\raise2pt\hbox{$<$}}}}}
\def\ga{\mathrel{\hbox{\rlap{\hbox{\lower4pt\hbox{$\sim$}}}{\raise2pt\hbox{$>$}}}}}

\def\d25{D$_{25}$}
\def\nh{{$N_{\rm H}$}}

\def\.25{0.25 keV\thinspace}

\def\mbh{\rm $M_{\rm BH}$}

\def\contin{\rm Continuum}

\def\hoix{\rm Holmberg\,IX X-1}

\shorttitle{Holmberg IX X-1: X-ray Outflows}
\shortauthors{D.~J. Walton et al.}

\begin{document}

\title{X-ray Outflows and Super-Eddington Accretion in the Ultraluminous X-ray
Source Holmberg IX X-1}

\author{D. J. Walton\altaffilmark{1},
J. M. Miller\altaffilmark{2},
F. A. Harrison\altaffilmark{1},
A. C. Fabian\altaffilmark{3},
T. P. Roberts\altaffilmark{4},
M. J. Middleton\altaffilmark{5},
R. C. Reis\altaffilmark{2}
}
\affil{
$^{1}$Space Radiation Laboratory, California Institute of Technology, Pasadena,
CA 91125, USA \\
$^{2}$Department of Astronomy, University of Michigan, 500 Church Street,
Ann Arbor, MI 48109, USA \\
$^{3}$Institute of Astronomy, University of Cambridge, Madingley Road,
Cambridge, CB3 0HA, UK \\
$^{4}$Department of Physics, Durham University, South Road, Durham DH1
3LE, UK \\
$^{5}$Astronomical Institute ‘Anton Pannekoek’, University of Amsterdam,
Postbus 94249, NL-1090 GE Amsterdam, the Netherlands
}

\begin{abstract}
Studies of X-ray continuum emission and flux variability have not conclusively
revealed the nature of ultra-luminous X-ray sources (ULXs) at the high-luminosity
end of the distribution (those with $L_{\rm X}\geq10^{40}$\,\ergps). These are
of particular interest because the luminosity requires either super-Eddington
accretion onto a black hole of $\sim$10\,\msun, or more standard accretion
onto an intermediate-mass black hole. Super-Eddington accretion models predict
strong outflowing winds, making atomic absorption lines a key diagnostic of the
nature of extreme ULXs. To search for such features, we have undertaken a long,
500\,ks observing campaign on \hoix\ with \suzaku. This is the most sensitive
dataset in the iron K bandpass for a bright, isolated ULX to date, yet we find no
statistically significant atomic features in either emission or absorption; any
undetected narrow features must have equivalent widths less than 15--20\,eV at
99\% confidence. These limits are far below the $\gtrsim$150\,eV lines expected
if observed trends between mass inflow and outflow rates extend into the
super-Eddington regime, and in fact rule out the line strengths observed from
disk winds in a variety of sub-Eddington black holes. We therefore cannot be
viewing the central regions of \hoix\ through any substantial column of material,
ruling out models of spherical super-Eddington accretion. If \hoix\ is a
super-Eddington source, any associated outflow must have an anisotropic
geometry. Finally, the lack of iron emission suggests that the stellar companion
cannot be launching a strong wind, and that \hoix\ must primarily accrete via
roche-lobe overflow.
\end{abstract}


\section{Introduction}

Ultraluminous X-ray sources(ULXs) are off-nuclear point sources found in nearby
galaxies that require extraordinary accretion-power. The nature of the most
luminous sources within this class -- those with $L_{\rm X}\geq10^{40}$\,\ergps\
(\eg \citealt{Farrell09, WaltonULXcat, Sutton12, Jonker12}) -- is particularly
interesting. These sources may be standard stellar-remnant black holes (\mbh\
$\sim$ 10\,\msun) accreting at super-Eddington rates (\citealt{Poutanen07},
\citealt{Finke07}), or intermediate-mass black holes (IMBHs: $10^{2}\lesssim$
\mbh\ $\lesssim10^{5}$\,\msun) accreting at sub-Eddington rates
(\citealt{Miller04, Strohmayer09a}). Indeed, the high-luminosity end of the ULX
distribution may include both extremes, or even a continuum in between
(\citealt{Zampieri09}). ULXs with $L_{\rm X}\geq10^{40}$\,\ergps\ represent a
regime in which our knowledge of black hole accretion can be extended and tested.
For recent reviews focusing on ULXs see \cite{Roberts07rev} and \cite{Feng11rev}.

A robust prediction for accretion at high rates (near Eddington or above) is that
strong outflows or winds should be launched from the accretion disk
(\citealt{Shakura73, Poutanen07, Ohsuga11, Dotan11, Vinokurov13}). Indeed,
Galactic stellar mass black holes (StMBHs) at moderately high accretion rates
(states dominated by thermal disk emission) frequently display evidence for such
disk winds (\citealt{Miller06a, Neilsen09, King12}), with outflow velocities
$v_{\rm out}\lesssim10,000$\,\kmps. When these outflows cover our line of
sight to the central source, absorption features are imprinted onto the intrinsic
X-ray continuum, the most prominent of which are typically the \ka\ transitions
of highly ionised iron (Fe {\small XXV} and/or {\small XXVI}). As expected, the
strength of the outflows observed appears to increase with the inferred accretion
rate in both StMBHs (\citealt{Ponti12}) \textit{and} in active galactic nuclei (AGN;
\citealt{King13}). For sub-Eddington StMBHs, outflows are seen predominantly in
high inclination sources, so the outflow geometry is inferred to be roughly
equatorial. Numerical simulations of winds from thin (sub-Eddington) disks
further support such an outflow geometry (\citealt{Proga04}).

The majority of ULXs have luminosities of a few $\times10^{39}$\,\ergps, and
likely represent a high luminosity extension of the disk-dominated thermal states
observed in Galactic StMBHs (\citealt{Kajava09, Middleton13nat}). Outflow
geometries in these cases are likely to still be largely equatorial. However, a
common prediction of super-Eddington accretion and the subsequent outflows
is that, as the accretion rate increases, the solid angle subtended by the outflow
should also increase (\citealt{Abramowicz05, King09, Dotan11}). At the high
Eddington rates required to explain the $L_{\rm X}\geq10^{40}$\,\ergps\ ULXs
(assuming \mbh\ $\sim10$\,\msun), one might expect that atomic iron features
associated with a strong, large solid angle outflow would be a common feature of
the X-ray spectra.

\cite{Walton12ulxFeK} describes initial searches for iron features in ULX spectra,
using archival \xmm\ data for two bright ($L_{\rm X}\sim10^{40}$\,\ergps)
sources, \hoix\ and NGC\,1313 X-1. No statistically significant features were
found in either source, and the limits obtained  required any lines to be relatively
weak in comparison to simple scaling of the sub-Eddington features observed
from other accreting black holes up to the super-Eddington regime. In order to
enhance the sensitivity to atomic iron features, we undertook deep observations
of the luminous source \hoix\ with \suzaku. In this Letter we present the results
from our search for X-ray spectral features in the iron-K energy range with this
new dataset.

\section{Data Reduction}
\label{sec_red}

\hoix\ was observed for a total exposure of $\sim$500\,ks during 2012 by the
\suzaku\ observatory (\citealt{SUZAKU}). To extract science products, we
reprocessed the unfiltered event files for each of the \xis\ CCDs (XIS0, 1, 3;
\citealt{SUZAKU_XIS}) and editing modes (3x3, 5x5) operational using the latest
\heasoft\ software package (version 6.13), as recommended in the \suzaku\ Data
Reduction Guide\footnote{http://heasarc.gsfc.nasa.gov/docs/suzaku/analysis/}.
Cleaned event files were generated by re-running the \suzaku\ pipeline with the
latest calibration, as well as the associated screening criteria files. For each of the
observation segments, source products were extracted with \xselect\ from circular
regions $\sim$200'' in radius, and the background was extracted from adjacent
regions free of any contaminating sources, with care taken to avoid the calibration
sources in the corners. Instrumental responses were generated for each individual
spectrum using the {\small XISRESP} script with a medium resolution. The spectra
and response files for the front-illuminated (FI) detectors (XIS0, 3) were combined
using the \ftool\ \addascaspec. Finally, we grouped the spectra to have a
minimum signal-to-noise (S/N) of 5 per energy bin with the {\small SPECGROUP}
task (part of the \xmm\ {\small SAS}), to allow the use of \chisq\ minimization
during spectral fitting.

\begin{figure}
\hspace*{-0.5cm}
\epsscale{1.12}
\plotone{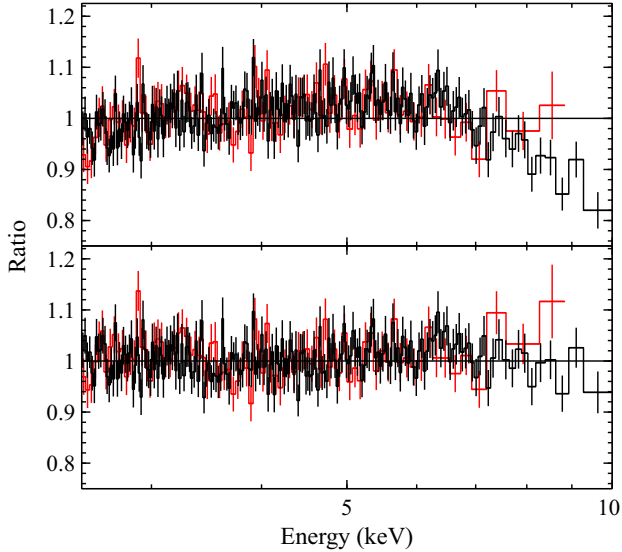}
\caption{
2.5-10.0\,\kev\ data/model ratios of the full \hoix\ \suzaku\ dataset to a simple
powerlaw ($\Gamma=1.72\pm0.01$; \textit{top panel}) and a powerlaw with a high
energy cut-off ($\Gamma=1.27\pm0.07$, $E_{\rm cut}=10^{+2}_{-1}$\,\kev;
\textit{bottom panel}). The latter provides a far superior fit to the data, indicating
a curved continuum is favoured.}
\label{fig_curv}
\end{figure}
\hspace{0.5cm}

\subsection{HXD PIN}

Due to the level of systematic uncertainty in the PIN background model
($\gtrsim$25\% of the `source' flux given the weak detection of the Holmberg\,IX
field; see \eg discussion in \citealt{Walton13spin}), and the variable nature of the
contaminating M\,81 nucleus (\citealt{Markoff08, JMiller10}), it is not possible to
constrain the high energy ($E>10$\,\kev) properties of \hoix\ with the collimating
PIN detector. The imaging capabilities of the recently launched \nustar\ observatory
(\citealt{nustar}) are required. Therefore, we do not consider the PIN data here, and
stress that any interpretation based on these data should be regarded with extreme
skepticism.

\section{Spectral Analysis}
\label{sec_spec}

Throughout this work, spectral modeling is performed with XSPEC v12.8.0
(\citealt{XSPEC}), and absorption by intervening neutral material is treated with
\tbnew\ (\citealt{tbabs}) using the appropriate solar abundances. We include two
absorption components, one fixed at the Galactic column ($N_{\rm H;
Gal}=5.54\times10^{20}$\,\atpcm; \citealt{nh}), and another with variable column
at the redshift of Holmberg\,IX ($z=0.000153$). During modelling, we only
consider data from the FI detectors in the 1--10\,\kev\ energy range, owing to a
calibration mismatch between XIS0 and XIS3 below $\sim$1\,\kev. For the same
reasons, we only consider data from the back-illuminated (BI) XIS1 detector over
the 2.5--9.0\,\kev\ energy range. We also exclude the 1.6--2.5\,\kev\ energy
range from the FI data owing to remaining calibration uncertainties associated
with the instrumental silicon K and gold M edges, and the 7.3--7.6\,\kev\ energy
range from the BI data owing to a residual background feature. The data from the
FI and BI detectors are modelled simultaneously, with all parameters tied between
the spectra, and we attempt to account for any further cross-calibration
uncertainties above 2.5\,\kev\ by allowing a variable multiplicative
cross-normalisation constant to vary between them. This value is always found
to be within $\sim$10\% of unity.

\subsection{Continuum Modelling}

We begin with a brief assessment of the form of the continuum, focusing first on
the 2.5--10.0\,\kev\ energy range. Specifically, we wish to determine whether a
simple powerlaw-like continuum is sufficient, or whether there is evidence for
curvature similar to other high quality ULX datasets (\citealt{Stobbart06,
Gladstone09, Walton4517}). Similar to previous works, we compare the results
obtained with a simple powerlaw continuum, with and without a high energy
exponential cut-off. We indeed find that allowing a cut-off offers a significant
improvement ($\Delta\chi^{2}=135$, one additional free parameter; see Fig.
\ref{fig_curv}), \ie the 2.5--10.0\,\kev\ spectrum does show curvature.

\begin{table}
\caption{\contin\ parameters obtained for \hoix.}
\begin{center}
\begin{tabular}{c c c}
\hline
\hline
\\[-0.2cm]
Component & Parameter & Value \\
\\[-0.3cm]
\hline
\hline
\\[-0.15cm]
\tbnew\ (Galactic) & \nh\ (cm$^{-2}$) & $5.54 \times 10^{20}$ (fixed) \\
\\[-0.225cm]
\tbnew\ (intrinsic) & \nh\ (cm$^{-2}$) & $(8 \pm 2) \times 10^{20}$ \\
\\[-0.225cm]
& $z$ &  0.000153 (fixed) \\
\\[-0.2cm]
\comptt\ & $T_{\rm seed}$ (\kev) & 0.1 (fixed) \\
\\[-0.2cm]
& $T_{\rm e}$ (\kev) & $2.6\pm0.1$ \\
\\[-0.2cm]
& $\tau$ & $7.1^{+0.3}_{-0.2}$ \\
\\[-0.2cm]
\hline
\\[-0.2cm]
\rchi & & 1730/1694 \\
\\[-0.25cm]
\hline
\hline
\end{tabular}
\label{tab_param}
\end{center}
\end{table}

Considering the 1--10\,\kev\ bandpass, we utilize a common parameterisation
of the continuum for ULXs, modeling the data with cool, optically thick
Comptonisation (\citealt{Stobbart06, Gladstone09, Middleton11b, Walton4517}).
We note that this parameterisation is largely empirical, and should not be
ascribed too much physical significance. The nature of the continuum emission
will be addressed in more detail in future work (Walton \etal \textit{in prep})
utilising the necessary \nustar\ data. Here, we use the \comptt\ code
(\citealt{comptt}), which provides an excellent fit (\rchi\ = 1730/1694). As we
do not consider the data below 1\,\kev, we are not sensitive to the presence of
any additional low-temperature ($\sim$0.2\,\kev) thermal component similar
to those seen previously in bright, unabsorbed ULXs (\citealt{Miller04}), so we
fix the seed photon temperature to 0.1\,\kev. The continuum parameters
obtained are summarised in Table \ref{tab_param}, and are broadly similar to
those obtained with previous observations (\citealt{Gladstone09, Vierdayanti10,
Walton12ulxFeK}).

\begin{figure}
\hspace*{-0.5cm}
\epsscale{1.12}
\plotone{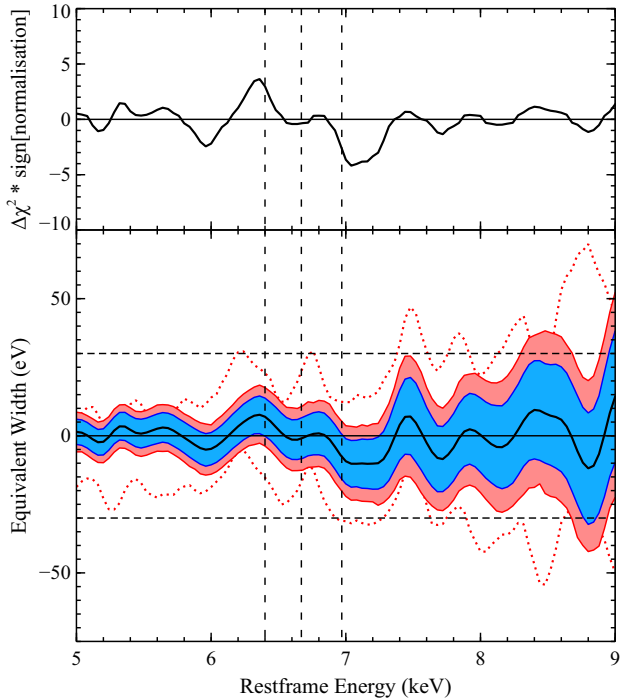}
\caption{
\textit{Top panel:} the $\Delta$\chisq\ improvement provided by a narrow
Gaussian line, as a function of (rest frame) line energy, for the full \suzaku\ \hoix\
dataset. Positive (negative) values of $\Delta$\chisq\ indicate the improvement is
obtained with an emission (absorption) line. We find no compelling evidence for
any narrow iron K features. \textit{Bottom panel:} 90\% (\textit{blue}) and 99\%
(\textit{red}) confidence contours for the equivalent width of a narrow line,
indicating the line strengths any narrow features intrinsically present could have
and still remain undetected. The rest frame transitions of neutral, helium-like and
hydrogen-like iron (6.4, 6.67 and 6.97\,\kev) are shown with vertical dashed lines,
and we also plot dashed horizontal lines representing $EW=\pm30$\,eV. Finally,
we also show the 99\% $EW$ limits obtained from archival \xmm\ data in
\cite{Walton12ulxFeK} for direct comparison (red dotted lines).}
\label{fig_limits}
\end{figure}

\subsection{Narrow Iron K Features}

\cite{Walton12ulxFeK} describes constraints on atomic iron K features for ULXs,
focusing on the archival \xmm\ data available for \hoix, revisited here with
\suzaku, and NGC\,1313 X-1. In order to search for atomic features in this new
dedicated dataset, we follow the same approach adopted in that work, varying a
narrow ($\sigma=10$\,eV) Gaussian across the 5--9\,\kev\ energy range, to
include the rest frame energies of the iron K transitions and also generously allow
for the possibility of blue-shifted features, in steps of 0.04\,\kev. For each energy
step, we record the $\Delta\chi^{2}$ improvement resulting from the inclusion of
the Gaussian line, as well as the best fit equivalent width ($EW$) and its 90 and
99\% confidence limits\footnote{We again note that we are computing confidence
limits on line strength, which are not formally the same as the upper limits on the
strength of lines that should be detected at a given confidence level
(\citealt{Kashyap10}), but are simpler to calculate, and provide more conservative
estimates for the maximum strength of any lines intrinsically present.}. The latter
quantities are obtained with the {\small EQWIDTH} command in \xspec, generating
10,000 parameter sets based on the best fit model and the covariance matrix,
which includes information on the model parameter uncertainties, and extracting
the confidence limits from the distribution of equivalent widths obtained.

The results obtained are shown in Fig. \ref{fig_limits}; the top panel shows the
$\Delta\chi^{2}$ improvement, multiplied by the sign of the best fit normalisation
to differentiate between emission and absorption, and the limits on $EW$ obtained
are shown in the bottom panel. For clarity, we highlight the energies of the \ka\
transitions of neutral, helium-like and hydrogen-like iron, as well as
$EW=\pm30$\,eV, representative of the strongest absorption features observed in
GRS\,1915+105 (\citealt{Neilsen09}). In addition, we also show the limits we
obtained previously with \xmm. As in our previous analysis, we find no statistically
significant line detections, so we again focus on the limits that can be placed
instead. Any narrow aromic features in the \suzaku\ data in the immediate Fe K
band (6--7\,\kev) must have equivalent widths less than $\sim$15--20\,eV (99\%
confidence); previously we were only able to  constrain such features to
$EW\lesssim30$\,eV. On average, the allowed $EW$ range at a given energy is a
factor of $\sim$1.5 smaller with our new dataset than obtained previously
(\citealt{Walton12ulxFeK}).

\section{Discussion}
\label{sec_dis}

Using our long 500\,ks \suzaku\ observation of \hoix, we have undertaken the
deepest study in the Fe K region of a bright, isolated ULX to date, to search for
iron absorption/emission features similar to those frequently observed in other
accreting black holes. Iron is ideally suited to identifying the presence of atomic
processes, as it is both difficult to fully ionize and cosmically abundant. Despite
the high sensitivity of the observations, we do not detect any discrete atomic iron 
features. The limits on any persistent narrow features (either in emission or
absorption) as yet undetected are now $EW\lesssim15-20$\,eV (99\%
confidence) over the immediate Fe K energy range.

\begin{figure}
\hspace*{-0.5cm}
\epsscale{1.12}
\plotone{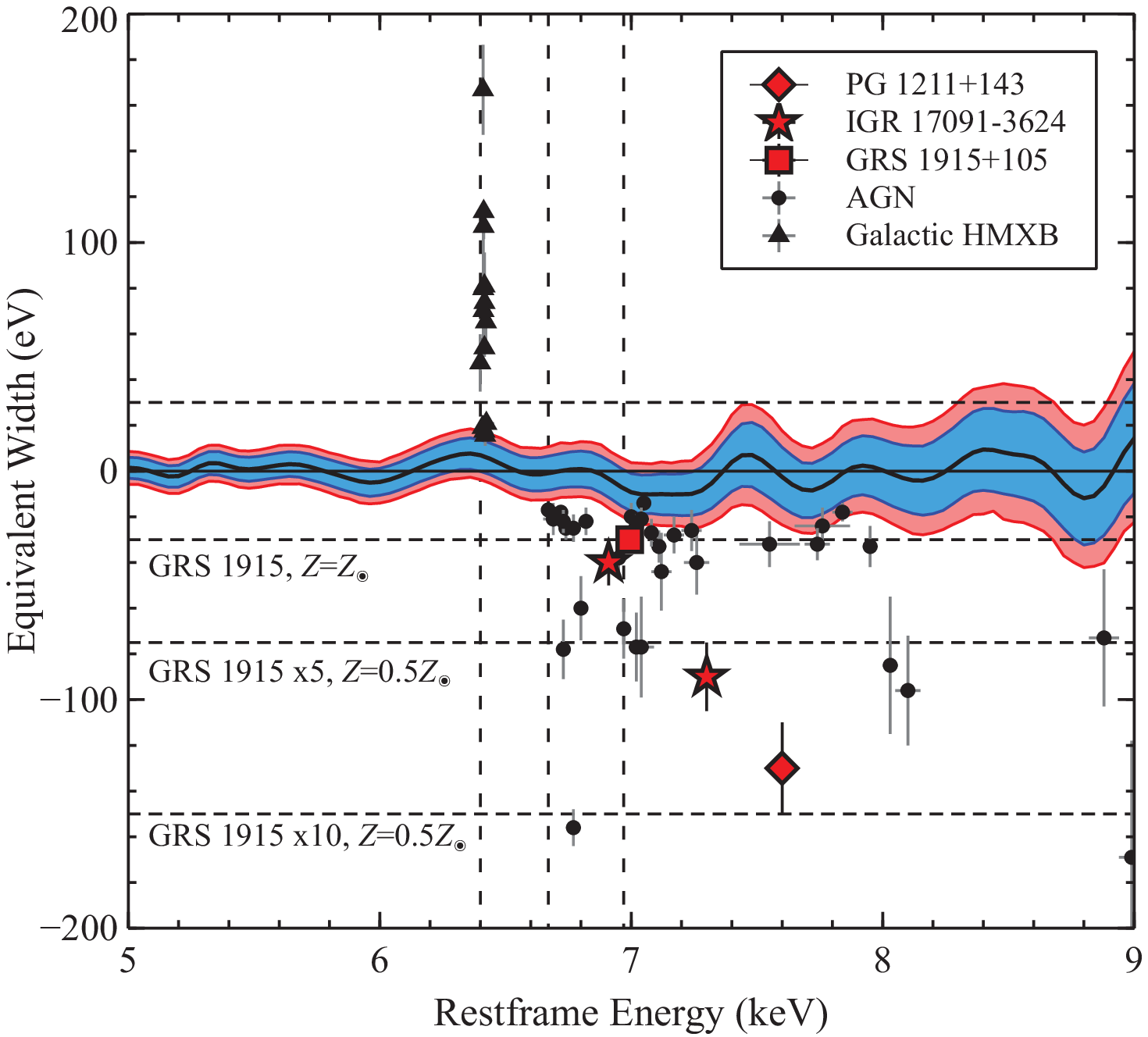}
\caption{
The line limits obtained for \hoix\ placed in the broader context of the
absorption features detected in both AGN (\citealt{Tombesi10b, Gofford13})
and selected Galactic StMBHs (\citealt{Neilsen09, King12}), and the emission
lines detected in Galactic HMXBs (\citealt{Torrejon10}). Some notable sources
have been highlighted. For further comparison, we also show the
GRS\,1915+105 absorption (assumed solar metallicity) scaled up by factors
of 5 and 10, accounting for the metallicity of Holmberg\,IX ($\sim$0.5\,\zsun;
\citealt{Makarova02}). These may be conservative scalings for the expected
$EW$, as the mass outflow rate should increase faster than linearly with
Eddington ratio.}
\label{fig_comp}
\hspace*{0.1cm}
\end{figure}

As shown in Fig. \ref{fig_comp}, these limits are now extremely restrictive when
considered in the context of the lines observed from other sources. Remarkably,
in terms of absorption, the constraints are such that any iron features present in
\hoix\ must be \textit{weaker} than the absorption lines observed in numerous
sub-Eddington Galactic StMBHs  (\citealt{Miller06a, Neilsen09, King12}) and AGN
(\citealt{Tombesi10b, Gofford13}). Meanwhile, any neutral iron emission from
\hoix\ must be weaker than that observed in the vast majority of Galactic
high-mass X-ray binaries (HMXBs), which ubiquitously display such emission
features (\citealt{Torrejon10}), and there cannot be any substantial emission
from helium- or hydrogen-like iron either. At the energies of these transitions,
the 99\% limits on emission lines are $EW<15$, 11 and 4\,eV respectively.

It is clear from the lack of absorption features that we cannot be viewing the
central hard X-ray emitting regions of \hoix\ through any substantial column of
partially ionized material. We can therefore exclude the presence of any massive,
spherical outflow in \hoix, as discussed in \cite{King03}, and wind-dominated
spectral models (\citealt{Vierdayanti10}) are also ruled out. If \hoix\ truly is a
StMBH accreting substantially in excess of the Eddington limit, we must be
viewing it through a line of sight that is not covered by the massive outflow
launched in such an accretion regime. The most natural conclusion is that, in
this scenario, we would have to be viewing the source close to face-on (see also
\citealt{Roberts11}), with the wind still retaining some kind of roughly equatorial
geometry, broadly similar to that inferred for sub-Eddington disk winds
(\citealt{Ponti12}), although presumably with a much larger scale-height for both
the disk and the outflow (\citealt{Poutanen07}). Furthermore, our results require
that any equatorial outflow must be launched further out than the central hard
X-ray emitting region, despite this presumably being the most irradiated region
of the accretion flow. If this is the case, observation of atomic iron emission from
such a massive outflow might be expected instead. The expected line profile
might well be more complex than a simple narrow Gaussian, however we stress
that there are no clear residual features that would readily be interpreted as
re-emission from a strong wind, which may prove problematic for any extreme
super-Eddington scenario. This will be addressed in more detail in future work.

The lack of narrow iron emission lines has strong implications for the broader
nature of the accretion onto \hoix, regardless of the black hole mass. In many
respects, ULXs are widely expected to be analogues to black hole HMXBs.
However, Galactic HMXBs ubiquitously display neutral iron emission lines
(\citealt{Torrejon10}), owing to illumination of the stellar winds in which they
are embedded, and from which they may primarily accrete. Since there is no
iron emission observed, either neutral or highly ionized, we infer that either the
stellar wind is weak (or even absent), or the accretion geometry is such that this
material is not illuminated by the hard X-rays emitted from the central regions.

Returning to the possible scenario in which ULXs are powered by a large scale
height, optically thick super-Eddington accretion flow; the inner regions of such
accretion flows are  expected to have a roughly conical geometry, with the
hard X-rays being produced in the hotter, central regions. If this funnel covers a
large enough solid angle, it may be able to shield the majority of the stellar wind
from the hard X-rays, which would be scattered and collimated preferentially out
of the equatorial plane. Given the expected large-scale extent of the stellar
wind, the optically thick regions of the accretion flow would have to
cover an extremely large solid angle in order to prevent substantial iron line
emission. However, the observed luminosity of the \heii\ $\lambda$4686
emission line, presumed to be associated with the cool outer disk, appears to
require that the X-ray luminosity cannot be highly anisotropic via photon
counting arguments (\citealt{Moon11}), and optical/UV photometry is also
indicative of X-ray reprocessing in the outer disk (\citealt{Grise11}), so it is not
clear that this offers a self-consistent scenario.

Alternatively, if the scale-height of the accretion flow is not sufficient to provide
substantial shielding, the lack of iron emission would therefore tell us that there
is no strong stellar wind. Indeed, for a spherically symmetric reprocessing
geometry, which works well for Galactic HMXBs (\citealt{Torrejon10}), and the
metallicity of Holmberg\,IX ($\sim$0.5\,\zsun; \citealt{Makarova02}), the
specific emission line limits obtained for neutral and hydrogen-like iron
correspond to reprocessing columns of \nh\ $\lesssim10$ and
$3\times10^{22}$\,\atpcm\ respectively (see \citealt{Walton12ulxFeK}). For
comparison, the HMXBs analysed by \cite{Torrejon10}, which have companion
masses \mstar\ $\gtrsim10$\,\msun, typically display columns of \nh\
$\gtrsim10^{23}$\,\atpcm. In turn, these observational constraints mean that
any stellar wind present most likely cannot provide the mass accretion rate of
\Mdot\ $\gtrsim1.5\times10^{-6}$\,\msun\ yr$^{-1}$ required to power the
observed X-ray luminosity from \hoix. Instead, \hoix\ most likely accretes via
Roche-lobe overflow, as suggested by \cite{Grise11}, the accretion mechanism
more typically associated with low mass X-ray binaries (LMXBs). However,
Galactic LMXBs are generally transient sources, spending the majority of the time
in quiescence, occasionally undergoing accretion events resulting in bright
$\sim$month-to-year long X-ray outbursts. In contrast, \hoix\ appears to be a
much more persistent source, requiring a sufficiently close binary system such
that Roche-lobe overflow/mass transfer remains roughly continuous.
\cite{Grise11} found a lower limit to the mass of the stellar companion of
\mstar\ $\gtrsim10$\,\msun. We suggest that the absent/weak stellar wind
implies that the companion cannot be substantially more massive than this
lower limit, given that more massive stars generally launch stronger winds.

Finally, in addition to improving the narrow line limits, we also confirm the
presence of curvature in the $\sim$3--10\,\kev\ energy range, as previously
indicated by \xmm\ (\citealt{Gladstone09}). We defer a detailed consideration
of the continuum to future work, which will focus on broadband spectral analysis
utilising the required \nustar\ data, although we do comment here that the
3--10\,\kev\ spectrum does not appear consistent with the powerlaw-like
emission expected from a standard sub-Eddington corona.


\section{Conclusions}

Our long \suzaku\ program on \hoix\ has provided the most sensitive dataset in
the Fe K region obtained for any luminous, isolated ULX to date. Despite the high
sensitivity of these data, we find no statistically significant narrow atomic features
in either emission or absorption across the 5--9\,\kev\ energy range. Furthermore,
the data are of sufficient quality to limit any undetected features to have equivalent
widths $EW\lesssim15-20$\,eV across the immediate Fe K bandpass at 99\%
confidence, \ie weaker than the features associated with sub-Eddington outflows
in a number of other black holes. Therefore, we cannot be viewing the central hard
X-ray emitting regions of \hoix\ through any substantial column of material.
Models of spherical super-Eddington accretion can be rejected, as can
wind-dominated spectral models. If \hoix\ is accreting at highly super-Eddington
rates, our viewing angle must be close to face on, such that the associated outflow
is directed away from our line-of-sight. Finally, the lack of iron emission implies
that the stellar companion is unlikely to be launching a strong wind, and therefore
the black hole must primarily accrete via roche-lobe overflow.

\section*{ACKNOWLEDGEMENTS}

The authors thank the reviewer for helpful comments. This research has made use
of data obtained from the \suzaku\ observatory, a collaborative mission between
the space agencies of Japan (JAXA) and the USA (NASA).

\bibliographystyle{/Users/dwalton/papers/mnras}

\bibliography{/Users/dwalton/papers/references}

\label{lastpage}

\end{document}